\documentstyle[multicol,prc,psfig,epsfig,aps
]{revtex}
\renewcommand{\(}{\left(}
\renewcommand{\)}{\right )}
\renewcommand{\[}{\left [}
\renewcommand{\]}{\right ]}
\def\fslash#1{#1 \!\!\! \slash}
\def\beq{\begin{equation}}
\def\eeq{\end{equation}}
\def\pa{\partial}

\def\varp{\varepsilon}
\def\bea{\arraycolsep .1em \begin{eqnarray}}
\def\eea{\end{eqnarray}}

\def\vp{{\bf p}}

\def\vk{{\bf k}}

\def\Tr{{\rm Tr}}

\let\si=\sigma

\let\om=\omega

\let\no=\nonumber

\def\eq#1{Eq. (\ref{#1})}
\def\refr#1{\cite{#1}}
\def\refrs#1{Refs.\cite{#1}}
\def\eqs#1{Eqs. (\ref{#1})}
\def\s0#1#2{\mbox{\small{$ \frac{#1}{#2} $}}}
\def\0#1#2{\frac{#1}{#2}}

\def\anp#1#2#3{Adv.\ Nucl.\ Phys. \ {\bf #1}, #2 (#3)}
\def\plb#1#2#3{Phys. Lett. {\bf B #1}, #2 (#3)}
\def\npa#1#2#3{Nucl. Phys. {\bf A #1}, #2 (#3)}

\def\prc#1#2#3{Phys. Rev.  {\bf C #1}, #2 (#3)}
\def\prd#1#2#3{Phys. Rev. {\bf D #1}, #2 (#3)} 
\def\prl#1#2#3{Phys. Rev. Lett. {\bf #1}, #2 (#3)}
\def\ann#1#2#3{Ann. Phys. (N.Y.) {\bf #1}, #2 (#3)}
\def\anp#1#2#3{Adv. Nucl. Phys. {\bf #1}, #2 (#3)}
\def\pr#1#2#3{Phys. Rep. {\bf #1}, #2 (#3)}

\def\ptp#1#2#3{Prog.\ Theor.\ Phys. \ {\bf #1}, #2 (#3)}

\def\ijmpe#1#2#3{Int.\ J.\ Mod.\ Phys.\ {\bf E #1}, #2 (#3)}
\def\jhep#1#2#3{J. High Energy Phys. \ {\bf #1}, #2 (#3)}
\begin{document}

\title{
In-medium meson effects on the equation of state
           of hot and dense nuclear matter}
\author{Ji-sheng Chen$^{1,2}\footnote{Email address: chenjs@iopp.ccnu.edu.cn}$~~~~~~Peng-fei
Zhuang$^{1}$~~~~~~
Jia-rong Li$^{2}$
}
\address{$^1$Physics Department, Tsinghua University, Beijing
100084,People's Republic of China}
\address{$^2$Institute of Particle Physics, Hua-Zhong Normal University, Wuhan 430079,
People's Republic of China}
\maketitle
\thispagestyle{empty}
\begin{abstract}
The influence of the in-medium mesons on the effective nucleon mass and in
	turn on the equation of state of hot/dense nuclear matter is discussed in
	the Walecka model. 
Due to the self-consistent treatment of couplings between
	nucleons and $\sigma $ and $\omega$ mesons, the temperature and density dependence
	of the effective hadron masses approaches more towards the Brown-Rho scaling
	law, and the compression modulus $K$ is reduced from $550$ MeV in mean field
	theory to an accepted value $318.2$~MeV.
\end{abstract}
\begin{multicols}{2}

The study of strongly interacting nuclear matter under extreme condition
	realized in relativistic heavy ion collisions has
	attracted a lot of attention during recent years(for example, see
	Ref. \refr{rapp2000} and references therein).
In high energy nuclear collisions the temperature may approach to
	several hundred MeV's or the density to several times of the normal nuclear
	density $\rho _0 $. 
It is believed that the property of hadrons in such high
	temperature or density will be
	quite different from that in vacuum. 
For instance, the behaviors of the in-medium
	hadronic masses may be qualitatively expressed by the so-called
	Brown-Rho scaling law\refr{brown1991}.

The quantum hadrodynamics QHD-I model (Walecka model ) and its extensions
	have been widely used to discuss the property of the symmetric nuclear matter
	and finite nuclei\refr{walecka1974,serot1986,serot1997,boguta1977,zm1990}. 
For example, the saturation properties at the normal density can be explained
	successfully by the mean field approximation.

In the Walecka model, the nucleons interact with each other through
	the exchange of $\sigma $ and $\om $ mesons. 
The $\si$ exchange
	gives the {\em attractive} force, while the $\om$ exchange the
	{\em repulsive} force. 
The mean field theory (MFT) has been widely
	used to discuss the nuclear matter saturation property at the
	normal density, while it was found that the obtained compression
	modulus $K\sim 550$ MeV is larger than the acceptable value
	$300\pm 50 $ MeV\refr{kouno1995}. 
The reason for so high compression
	modulus is that the effective nucleon mass drops too fast with the
	density. 
By taking into account the vacuum fluctuations, the
	behavior of the effective nucleon mass with density may be cured
	to some extent, but the obtained compression modulus $K$ $\sim
	450$ MeV is still larger than the acceptable value. To solve this
	problem, the $\sigma-\om $ model including nonlinear interaction
	terms $b\sigma ^3 +c\sigma ^4 $ of $\si$ and the Zimanyi-Moszkowski
	(ZM) model have been proposed\refr{boguta1977,zm1990}. 
With more adjustable parameters
	in the nonlinear $\si -\om $ version, the effective nucleon mass
	drops slowly and the compression modulus $K$ may be in the region
	of ``appropriate values." 
The behaviors of the effective nucleon mass and the compression
	modulus $K$ may be also cured to some extent with the nonrenormalizable 
	ZM model\refr{zm1990}.

On the other hand, the in-medium effects of hadrons under
	extreme environment have been emphasized in the QHD-like framework. 
Since Walecka model's Lagrangian is a truncated form of a chirally symmetric
	Lagrangian with the vector and scalar fields taken as
	chiral-singlet fields\refr{graciela1995,brown1996}, it is argued
	that the predictions about the effective masses of hadrons in
	hot/dense environments by QHD are intrinsically consistent with
	the chiral symmetry restoration\refr{aguirre2001,cohen1991}.
This simple model includes implicitly vacuum effects and subnucleon
	structures.
With QHD-I and by considering the Dirac sea
	contribution, the obtained effective masses of light vector mesons
	drop down with increasing density/temperature. 
However, in discussing the hadronic masses, the relevant equations of QHD-I
	model had not been solved with a consistent manner, i.e., first
	solving the equations for the effective nucleon mass $M_N^*$ and
	chemical potential $\mu _N^*$ (for finite temperature occasion) in
	MFT or RHA (relativistic Hartree approximation) approach and then using the obtained $M_N^*$ and $\mu
	^*_N$ to get the full propagators for mesons $\sigma $ and $\om $
	and even $\rho $\refr{shiomi1994,jean1994,dutt1997,Iwasaki2000}.
Therefore, the in-medium meson effects were not reflected in the
	effective nucleon mass  as pointed out by Bhattacharyya {\it et
al.} in
	Ref.\refr{Bhattacharyya1999}. 
Our motivation is to study the effect of
	nonperturbative  in-medium modification of $\si $ and $\om $ on the
	effective nucleon mass and on the equation of state (EOS) of hot/dense nuclear matter
	by using renormalizable original version QHD-I. 
With Dyson-Schwinger Green
	function approach, the in-medium resummed nucleon and meson propagators are
	treated self-consistently.

\begin{figure}[ht]
	\centering
	\psfig{file=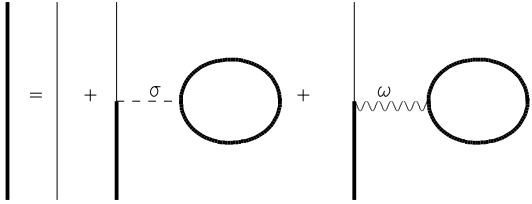,width=7.cm,angle=-0}~\\[.2cm]
	\caption{
		\small
		Diagrammatic representation for the nucleon propagator in relativistic
		Hartree approximation (RHA).
	}\label{fig1}
\end{figure}

The full description for the Lagrangian of Walecka model can be found in
	Refs.\refr{serot1986,chen2002}.
Under the mean field approximation, the full nucleon propagator in the
	medium as indicated in Fig.\ref{fig1} is attributed to the calculation of the
	tadpole self-energies with finite temperature field theory:
\bea
&&\Sigma _s=-\0{g_\si^2}{m_\si^2}T\sum _{p_0}
\int _\vp
\Tr \0{1}{{\fslash{p}}-M_N^*},
\\
&&\Sigma _v =-\0{g_\om ^2 }{m_\om ^2 }\gamma ^\mu T \sum
_{p_0}  \int _\vp
\Tr \gamma _\mu \0{1}{{\fslash{p}}
-M_N^*},
\eea
where ${p_0}=(2 n +1 ) \pi T i +\mu _N^*$  with $T$ and $\mu ^*_N$ being
	temperature and effective baryon chemical potential, respectively, and the symbol $\int
	_\vp=\int\0{d^3\vp }{(2 \pi )^3}$.  
With the residue theorem, one
	can divide the self-energies into the vacuum fluctuation and
	the matter(obviously related to the distribution functions) parts, and define
	 the effective nucleon mass $M_N^*$ and effective chemical potential $\mu_N^*$\refr{serot1986},
\bea\label{self1}
M^*-M=&&-\0{\gamma g_\si^2 }{m_\si^2}
 \int _\vp
 \0{M_N^*}{E_N^{*}}
(n_N +{\bar n}_N
)
+\Delta M_{vac}^*,
\\
\label{self2}
\mu_N^*-\mu_N =&&-\0{g_\om ^2 }{m_\om ^2 }\rho _B,
\eea
where the baryon density
\bea
\rho _B =\gamma\int _\vp
(n_N
-{\bar n}_N),
\eea
with
\bea
n_N
=&&\0{1}{e^{\beta (E_N^*
 -\mu _N^* )}+1},~~~~~
{\bar n}_N
=\0{1}{e^{\beta (E_N^* +\mu _N^* )}+1},\no\\
E_N^{*}=&&\sqrt{p^2+M_N^{*2}}.\no
\eea
The spin-isospin degenerate factor is  $\gamma=4$ for symmetric nuclear matter
	and $2 $ for neutron star.
One can see that the self-consistent equation of effective nucleon mass in RHA is just
	the result
	in the MFT plus the vacuum fluctuation contribution $\Delta
	M_{vac}^*$. 
Similarly,
	the energy density and the pressure are the MFT results plus
	the vacuum fluctuation contribution $\Delta \varp _{vac}^*$,
\bea\label{density}
\varp=&&\0{m_\sigma ^2 }{2 g_\sigma ^2 } (M_N-M_N^*)^2+\0{g_\om ^2}{2 m_\om
^2}\rho _B^2+\no\\
&&~\gamma \int _\vp E_N^*
(n_N+{\bar n_N} )-\Delta
\varp _{vac}^*
;\\\label{pressure}
P=&&-\0{m_\sigma ^2 }{2 g_\sigma ^2 } (M_N-M_N^*)^2+\0{g_\om ^2}{2 m_\om
^2}\rho _B^2-\no\\
&&~ \gamma T \int _\vp
\[\ln (1-n_N)+\ln (1-{\bar
n_N} )\]+\Delta \varp _{vac}^*+B^*.
\eea
In \eqs{self1}, (\ref{density}) and (\ref{pressure}),
	the explicit expressions for the
	vacuum fluctuation contributions $\Delta
	M_{vac}^*$ and $\Delta\varp _{vac}^*$ can be found in \refrs{serot1986,chin1977}.
The thermodynamics compensatory term $B^*$ in \eq{pressure} can be determined uniquely by the
	thermodynamics self-consistency relation between the energy density $\varp $ and pressure $p$.
	For MFT and RHA approaches, this term is vanishing.

\begin{figure}
	\centering
	\psfig{file=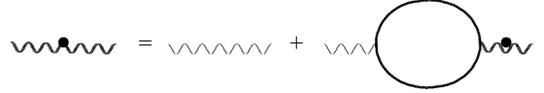,width=7. cm,angle=-0}~\\[.2cm]
	\caption{ \small The in-medium $\om $ propagator in random phase
		approximation.
		}\label{fig2}
\end{figure}

The properties of mesons are studied by the meson propagators in medium, which
	are normally calculated by using the random phase approximation (RPA) with the full
	nucleon propagator in Fig.\ref{fig1}.
The Dyson-Schwinger equation of the $\om $ meson propagator is indicated in
	Fig.\ref{fig2} and its solution for the full propagator $D ^{\mu\nu }$
	can be determined by the polarization tensor $\Pi ^{\mu\nu}(k)$,
\bea
	\Pi ^{\mu\nu }(k) =(D^{-1})^{\mu\nu}-(D_{(0)}^{-1})^{\mu\nu},
\eea
where $D^{\mu\nu}_{(0)}$ is the bare propagator. Using the Feynman rules at
	finite temperature\refr{kapusta1989}, one has
\bea
	\Pi ^{\mu\nu}(k) =&&2 g^2_{\om} T \sum _{p_0}
	\int _\vp
	 \Tr \[\gamma ^\mu \0{1}{{\fslash{p}}-M_N^*}
	\gamma ^\nu
	\0{1}{{\fslash{p}} -\fslash{k}-M_N^*} \].
\eea

Different from the occasion in vacuum,
	there are two independent elements $\Pi ^{L (T)} $ for the polarization tensor
	$\Pi ^{\mu\nu }(k)$ in medium,
\bea
	\Pi ^{\mu\nu}(k) =&&\Pi ^L (k) P_L ^{\mu\nu} +\Pi ^T (k) P^{\mu\nu }_T,\no\\
	D^{\mu\nu}=&&-\0{P_L^{\mu\nu}}{k^2 -m_\om ^2
	-\Pi_L(k)}\no\\ &&-\0{P_T^{\mu\nu}}{k^2
	-m_\om ^2 -\Pi_T(k)}-\0{k^\mu k^\nu}{m^2_\om k^2},
\eea
where the $P_L ^{\mu\nu}$ and $ P^{\mu\nu }_T$ are the standard projection tensors \refr{kapusta1989}.
The ingredients $\Pi_L(k) $ and  $\Pi_T(k) $ are determined by
\bea
	\Pi_L(k)=\0{k^2}{\vk ^2 }\Pi ^{00}(k),~~~~~
	\Pi_T(k)=\012 P^{ij}_T\Pi_{ij}(k).
\eea
The pole position of the full propagator $D^{\mu\nu}$  determines completely
	the dispersion relation of $\om$ meson excitations in medium. The expressions for the
	various components of $\Pi ^{\mu\nu }(k)$ are similar to those of $\rho $ 
	as given in Refs.\refr{chen2002,chen20029}
	except the vanishing of the tensor coupling constant for $\omega $
meson, i.e., $\kappa _\om =0$. 

Analogously to the discussion of $\om $ meson, one can discuss the
	property of $\si$ in the medium. Its diagrammatic Dyson-Schwinger equation
	is similar to Fig. \ref{fig2} and the self-energy
	is\refr{chen20029,caillon1993}
\bea
	\Pi (k)=&&2 g_\si ^2 T \sum _{ p0} \int _\vp
	\Tr \[\0{1}{{\fslash{p}}-M_N^*}
	\0{1}{{\fslash{p}}-\fslash{k}
	-M_N^*}\]\no\\
	=&&\0{3 g_{\sigma}^2}{2 \pi^2} \left [3(M_N^{*^2} -M_N^2) -
	4(M_N^*- M_N) M_N\no\right. \\
	&&\left.
	- (M_N^{*^2} - M_N^2) \int_{0}^{1}
	\ln\frac{M_N^{*^2}- x(1 - x)k^2}{M_N^2} dx \no\right. \\
	&&\left.- \int_{0}^{1} \(M_N^2 - x(1 - x)k^2\) 
	\ln\frac{M_N^{*^2} - x(1 -x)k^2}{M_N^2 - x(1 - x)k^2} dx\right ]
	\no\\&&
	+\0{g_\sigma ^2}{\pi ^2}\int \0{p^2 dp }{E_N^*} (n_N+{\bar n_N})
	\[2+\0{k^2-4 {M^*_N}^2  }{4 p |\vk|}(a+b )\],\no\\
\eea
where
\bea
	a=&&\ln\0{k^2-2 p |\vk|-2 k_0 E_N^* }{k^2+2 p |\vk|-2 k_0 E_N^*},~~~
	b=\ln\0{k^2-2 p |\vk|+2 k_0  E_N^*}{k^2+2 p |\vk|+2 k_0  E_N^*}.\no
\eea

There are two kinds of effective meson masses 
	related to our work.
One is the pole mass $m^*$, which can be matched to the results obtained by
such as QCD sum rules. It is defined by
	the pole position of the meson
	propagator in medium by taking the limit $|\vk|\rightarrow 0$ of $\Pi (k)$,
\bea\label{effmass}
	k_0^2 -m ^2 -\lim _{|\vk | \rightarrow 0}\Pi  (k)=0.
\eea
The other  is the off-shell mass ${\bar m} $  in the medium determined by
\bea\label{self3}
	{\bar m}^{2}=m^2 +\lim_{k_0\rightarrow 0}\lim_{|\vk | \rightarrow 0}\Pi (k).
\eea
Due to the fact that the longitudinal and transverse components of the $\om $
	polarization tensor $\Pi _\om ^{\mu\nu} (k)$ coincide with each other in taking the limit
$|\vk | \rightarrow 0$, 
	the subscripts $L$ and $T$ for $\Pi _\om (k)$ are omitted.

As calculated in the following, the meson property will be quite different
	from the vacuum scenario. 
To consider the back interactions of $\si$ and $\om$ mesons with nucleon $N$
	from the point of view of self-consistency, the {\it in-medium} $\sigma$ and
	$\omega $ propagators with vanishing four-momentum transfer should be used in
	determining the nucleon propagator as indicated by Fig.\ref{fig1} due to the
	adapted Hartree approximation.
Therefore, the meson masses $m_\si $ and $m_\om $ in the
	effective nucleon mass and chemical potential equations
	(\ref{self1}) and (\ref{self2}),
	and in the EOS's (\ref{density}) and (\ref{pressure})
	should be the off-shell meson masses ${\bar m} $'s determined by \eqs{self3}. 
It means that \eqs{self1} and (\ref{self2})
	obtained by the RHA of the nucleon propagator and \eqs{self3} obtained
	by the RPA of the meson propagator form a closed set of equations,
	and should be solved simultaneously.
\begin{figure}[h]
\centering
\psfig{file=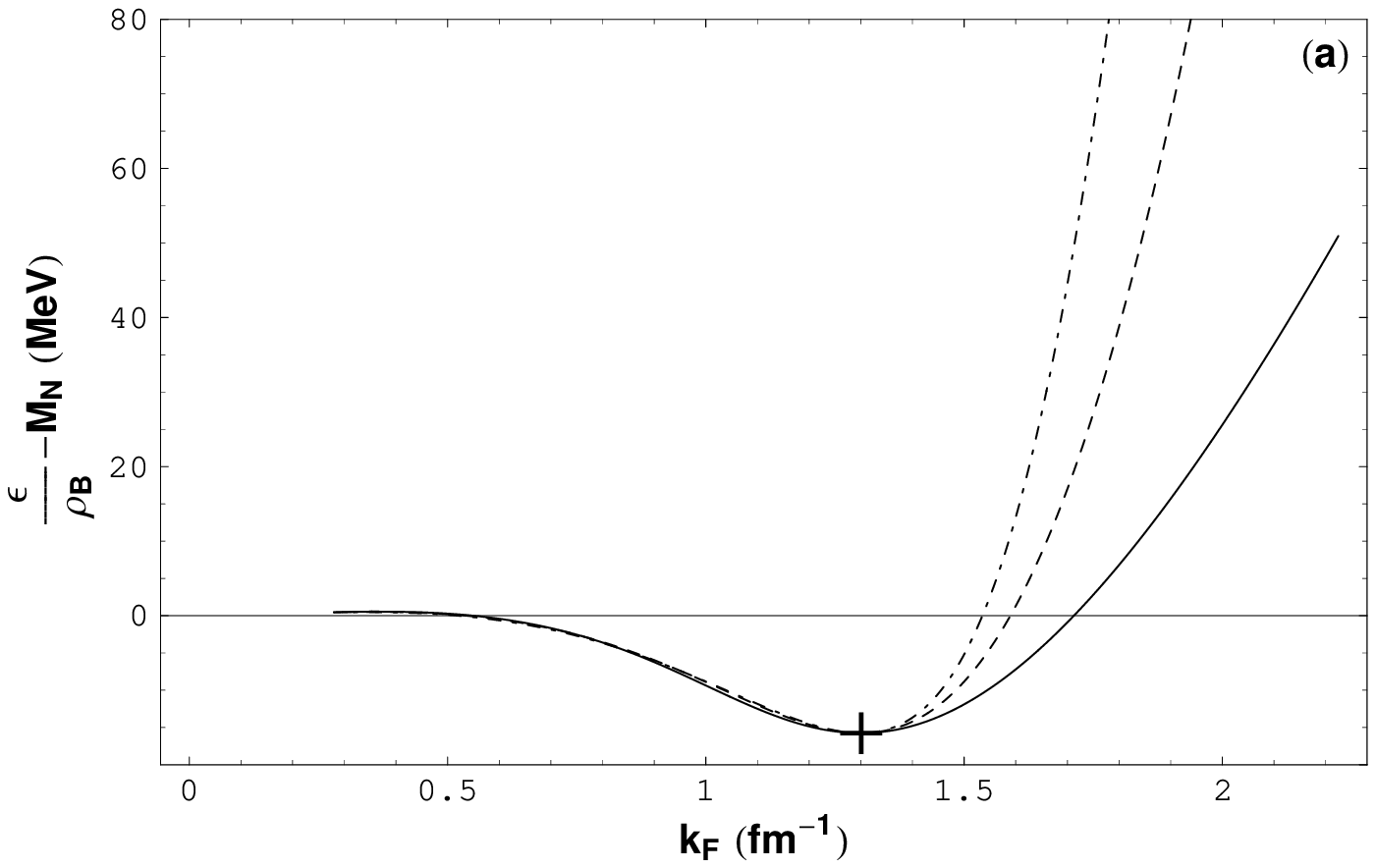,width=7.cm,angle=-0}\\[.2cm]
\psfig{file=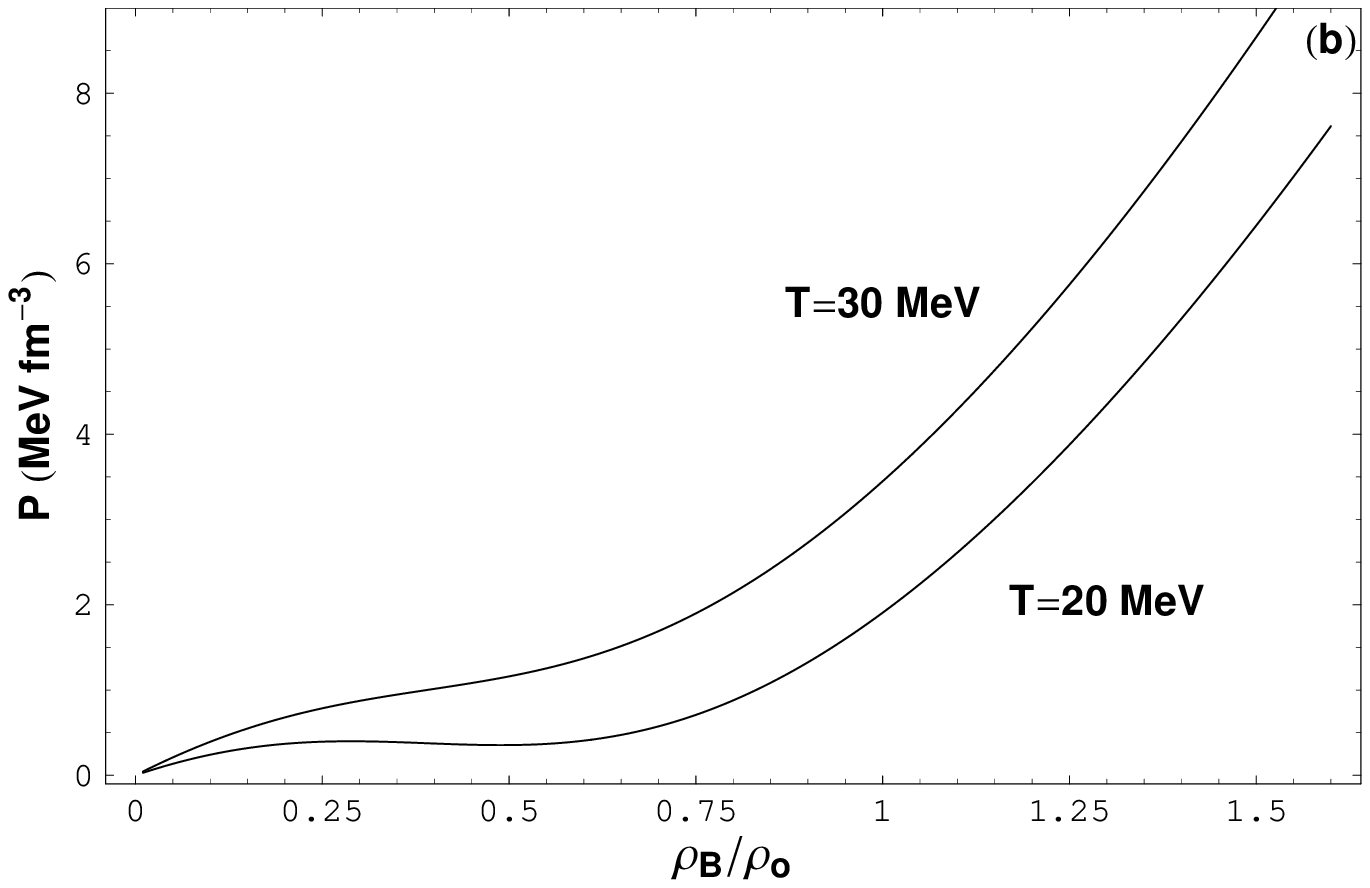,width=7.cm,angle=-0}
~\\[.2cm]
 \caption{(a) The binding energy for nuclear matter as a function of Fermi momentum at $T=0$.
The dot-dashed, dashed, and solid
lines represent the MFT, RHA, and our results, respectively;
(b) The pressure as a function of rescaled density $\rho _B/\rho_0$ for
two temperatures $T=20$ MeV  and $30$ MeV.}\label{fig3}
\end{figure}

Considering the in-medium effects of the off-shell masses (\ref{self3}) on the EOS, 
	one must at first refit the binding energy at the normal nuclear density at zero  temperature 
	by adjusting the coupling constants $g_{\om (\si)} $  as done in MFT and RHA.
The coupling constants are listed in Table \ref{tab}, 
	which are smaller  than those of MFT and RHA. 
It is interesting to note that the EOS
	becomes much softer and the compression modulus
\bea\label{compressibility} K=9 \rho _B^2 \0{\pa^2 e_N}{\pa \rho^2
_B}\Huge{|}_{\rho _B=\rho _0} 
\eea 
is reduced to $K=318.2$ MeV, which is acceptable in dealing with realistic nuclear matter.
In \eq{compressibility}, $e_N$ is the binding energy ($\varp /\rho _B-M_N$). 
Of course, this interesting numerical result can be also obtained with the nonlinear $\sigma-\omega$
version by adjusting the additional parameters due to the renormalizability
of original QHD-I as pointed out in the introduction\refr{serot1997}.
However, it is interesting enough that it has been obtained here from the point of view of in-medium meson
contribution on the nucleon and bulk property of nuclear matter.  
To the best of our knowledge, a similar numerical result of compression modulus $K$ has also been obtained long ago by Ji 
	in Ref. \refr{ji1988} with QHD-I 
	through taking into account the polarization effects at zero
temperature. 
It should be pointed out that our numerical results (especially $g_\si $, $g_\om
$, and $K$) are
	also different from those of Ref.\refr{Bhattacharyya1999} for the
	zero-temperature scenario.

One can further study the energy density and pressure.
One should note that the consideration of the back interactions of
	in-medium mesons with nucleons will lead to the nonvanishing compensatory term $B^*$
	in \eq{pressure} due to the implicit higher order contribution from the
	resummed meson propagators instead of frozen ones. 
In principle, this term can be expanded into a series of $M_N^*-M_N$. 
With the energy density \eq{density} and thermodynamics
	self-consistency condition at zero temperature
$p=\rho_B ^2 \(\pa/\pa \rho_B\)\(\varp /\rho_B\)$,
this compensatory term can be fixed uniquely. 
In the upper panel (a) of Fig.\ref{fig3}, we give out the binding energy at $T=0$, from which one can confirm
 that the equation of state really
becomes softer and the  compression modulus is much smaller than the results
of MFT and RHA.
The first order liquid-gas phase transition in the
low temperature case still exists as indicated by 
Fig.\ref{fig3}(b) and the
critical temperature $T_c$ is about $21$ MeV.
\begin{figure}[h]
 \centering
\psfig{file=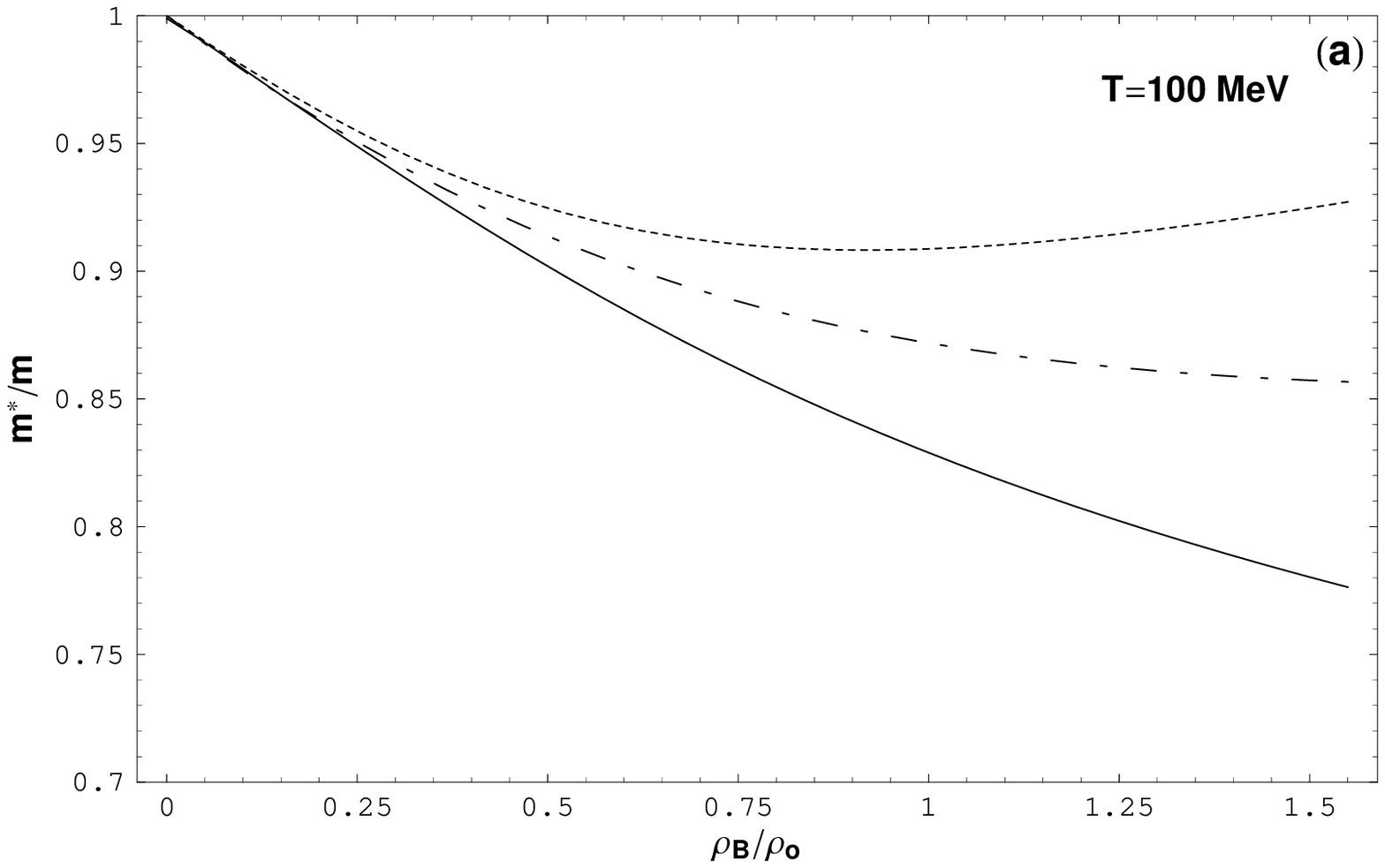,width=7.cm,angle=-0}~\\[.2cm]
\psfig{file=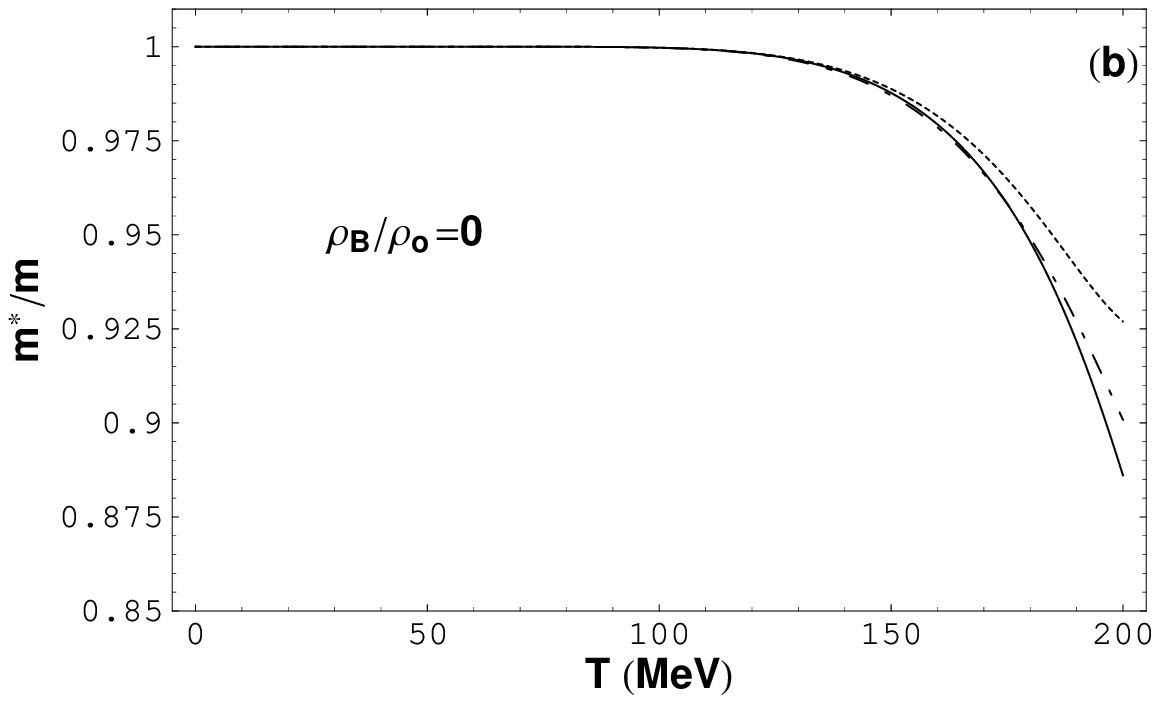,width=7.cm,angle=-0}
~\\[.2cm]
 \caption{
\small
The effective masses $M_N^*$ (solid lines ), $m_\si ^*$ (dotted lines) and $m_\om ^*$
(dot-dashed lines) as functions of rescaled density $\rho _B/\rho_0$ at
$T=100$ MeV (a) and temperature $T$ at $\rho _B=0$ (b), respectively.}\label{fig4}
\end{figure}

The pole masses $M_N^*$, $m_\si ^* $ and $m_\om ^* $  determined
by \eqs{self1} and (\ref{effmass}) are indicated in Fig.\ref{fig4}.
Because of considering the back interactions of $\si $ and $\om$
mesons with nucleons, their temperature and density dependence
approaches more towards the Brown-Rho scaling law compared with
the results of MFT/RHA. The key point is that the exchanged mesons
are not bare but in-medium ones, their masses are
self-consistently determined by the Dyson-Schwinger equations. The
temperature and density dependence of hadron masses studied here
is analogous to that of hidden local symmetry theory of
Harada and Yamawaki where such dependence is required by the
Wilsonian matching to QCD\refr{harada2001}. It is interesting to
note that our results are consistent with those obtained in recent
works\refr{brown2002,song2001}, where the Brown-Rho scaling fits
naturally into the relevant framework. This consistency is
attributed to the hidden chiral symmetry in QHD-I and should contribute to
the understanding of hadronic matter under extreme conditions within the relativistic nuclear theory
framework.

In summary, we have investigated the in-medium meson contribution
to the EOS of hot/dense nuclear matter with the original Walecka
model by treating the coupled nucleon and meson propagators
self-consistently. This kind of
contribution makes the EOS softer than the RHA and MFT approaches.
Due to the couplings between nucleon and in-medium mesons, the
compression modulus $K$ drops from about $550$ MeV to about $320$
MeV and the effective hadronic  masses approach more towards the
Brown-Rho scaling law, and the low temperature liquid-gas phase
transition still exists.

{\bf Acknowledgements}:
Beneficial communications and discussions with Professor S. Phatak,
Professor S. K. Ghosh, 
and Professor Zh.-X Li are acknowledged by one of the authors (J.-S Chen). 
This work was supported
by the NSFC under Grants Nos. 10135030, 10175026, 19925519 and the China
Postdoc Research Fund.
\end{multicols}

\begin{table}
\caption{
	The coupling constants $g_\si$ and $g_\om $, the effective nucleon mass
	$M_N^*$ and the compression modulus $K$ at normal nuclear density $\rho _0
	=0.1484 $fm$^{-3}$ at $T=0$. In MFT, RHA, and our approach(labeled as RHA+RPA),
	the nucleon and $\om$ meson masses in vacuum are taken to be $M_N=939$ MeV,
	$m_\om =783$ MeV, and the $\si $ meson mass in the vacuum is selected to be
	$m_\si =520$ MeV (in the literature with MFT) and $458$ MeV. To compare with the results in
	MFT and RHA, we have also shown the {\it medium dependent} coupling constants $C_s^2
	=g_\si^2(M_N^2/\bar m_\si^2)$ and $C_v^2
	=g_\om^2( M_N^2/\bar m_\om^2)$.}
   \begin{tabular}{cccccccc}
        & $g_\sigma^2$    &   $g_\om^2$  & $m_\sigma $ (MeV) & $C_s ^2$&$C_v^2$& $K ~$ (MeV) &$\0{M_N^*}{M_N}$ \\
     \tableline
        MFT &109.94&191.04 &520 &358.49 &274.76&547.2&0.540\\
	&85.286&191.04 &458&358.49 &274.76&547.2&0.540 \\
     \tableline
        RHA & 54.289&102.76 & 458 & 228.19&147.78&452.6&0.731  \\
     \tableline
        RHA+RPA &42.229 &69.729&458&152.59 &86.263&318.2&0.803 \\
   \end{tabular}
 \label{tab}
\end{table}
\begin{multicols}{2}

\end{multicols}
\end{document}